# The impact of economic policy uncertainties on the volatility of European carbon market

Peng-Fei Dai [a, b *], Xiong Xiong [b], Toan Luu Duc Huynh[c, d, e], Jiqiang Wang [b *]

[a] *School of Business, East China University of Science and Technology, Shanghai, 200237, China.*
[b] *College of Management and Economics, Tianjin University, Tianjin 300072, China.*
[c] *Chair of Behavioral Finance, WHU-Otto Beisheim School of Management, Vallendar, Germany.*
[d] *IPAG Business School, Paris, France*
[e] *University of Economics Ho Chi Minh City, Ho Chi Minh City, Viet Nam*

## Abstract

The European Union Emission Trading Scheme is a carbon emission allowance trading system designed by Europe to achieve emission reduction targets. The amount of carbon emission caused by production activities is closely related to the socio-economic environment. Therefore, from the perspective of economic policy uncertainty, this article constructs the GARCH-MIDAS-EUEPU and GARCH-MIDAS-GEPU models for investigating the impact of European and global economic policy uncertainty on carbon price fluctuations. The results show that both European and global economic policy uncertainty will exacerbate the long-term volatility of European carbon spot return, with the latter having a stronger impact when the change is the same. Moreover, the volatility of the European carbon spot return can be forecasted better by the predictor, global economic policy uncertainty. This research can provide some implications for market managers in grasping carbon market trends and helping participants control the risk of fluctuations in carbon allowances.

**Nomenclature**

EU: European Union

EU ETS: European Union Emissions Trading System

RCP: return series of the carbon price

RV: realized volatility of the carbon price returns

EPU: economic policy uncertainty

EUEPU: European economic policy uncertainty

GEPU: global economic policy uncertainty

PCA: principal component analysis

RCP: natural logarithmic return series of the carbon p rice

1. Introduction

To cope with global climate change, in December 1997 in Kyoto, Japan, the Kyoto Protocol was formulated by countries participating in the United Nations Framework Convention on Climate Change, and its goal is to control greenhouse gas emissions in the atmosphere to a certain level to prevent climate change from causing further harm. Developed countries have undertaken the obligation to reduce carbon emissions since 2005, while developing countries began undertaking this obligation in 2012. Against this background, European Union (EU) member states formally ratified the Kyoto Protocol on May 31, 2002. To achieve its carbon emission reduction goals, the EU took the lead in establishing the European Union Emissions Trading System (EU ETS), which provides empirical inspiration for the establishment of other carbon markets. This is currently the largest carbon trading system in the world and has thus far reached the third stage. Phase I, also known as the trial operation phase, extended from 2005 to 2007 and Phase II from 2008 to 2012, while Phase III was from 2013 to 2020.

Since the establishment of the EU ETS, many scholars have analysed the impact of the scheme on the economy. This type of research largely focuses on the impact on the energy industry. Amongst human activities, the burning of fossil fuels is considered as the factor which most affects climate change, accounting for 80% of total human greenhouse gas emissions. Fossil fuels are the main power generation input in the energy industry, which generates considerable carbon emissions. Therefore, the establishment of the EU ETS has a strong theoretical influence on the energy industry. Numerous studies have indicated that the carbon market has brought new impacts to the power industry, especially with respect to the costs to companies of power generation (Delarue et al., 2007). The EU ETS changes the marginal cost of power generation and thus also affects the electricity prices (Kara et al., 2008; Zachmann and Hirschhausen, 2008), affecting the profits of power generation companies as well as investment in emission reduction (Bonenti et al., 2013). With the progressive substitution of new energy for traditional energy, many researchers find that the establishment of the carbon market play an indirect and strong role in promoting the development of the new energy industry (Hobbie et al., 2019).

Some scholars have also studied the impact of the system on non-energy industries with energy-intensive characteristics, primarily the aviation industry and some industrial production.

Efthymiou and Papatheodorou (2019) find that the inclusion of the aviation industry in the emission trade system has not enabled this industry to achieve significant emission reductions. The previous literature reports that the inclusion in the emission trade system do not produce a negative impact on the development of the entire aviation industry (Zhang and Wei, 2010). Some literatures (Asselt and Biermann, 2007; Oberndorfer and Rennings, 2010) investigate the influence of the EU ETS on the short-term competitiveness of EU companies in the cement, steelmaking, steel and aluminium industries, suggesting that the EU ETS has also had little effect on them. However, the researchers do believe that as the market environment changes in the future, such influence may increase. Other scholars have analysed the effect of EU ETS from the integral perspective of related enterprises (Löschel et al., 2019; Lutz, 2016; Calel and Dechezlepretre, 2016). Their empirical results show that the establishment of the carbon market promotes the enterprises' low-carbon production (Löschel et al., 2019; Lutz, 2016) and investment in low-carbon technologies (Calel and Dechezlepretre, 2016). Brouwers et al. (2016) provides the evidence that the occurrence of the EU ETS has even influenced the market value of regulated enterprises. In a word, the EU ETS is closely related to the entire socio-economy.

Interestingly, not only the carbon market affects the social economy, but also the carbon market is affected by the social economy in reverse. Carbon prices depend largely on coal, oil, natural gas and electricity prices (Chevallier, 2011; Aatola et al., 2013; Creti et al., 2012). Some studies have shown that rising energy prices will promote higher carbon prices, while the falling energy prices will cause lower carbon prices (Zhang and Sun, 2016). However, with the operation and adjustment of the carbon market, the fundamental factors affecting the carbon price also changed. Especially in Phase I, the carbon price occurs a sharp decline, which leads to the structural breakpoint in the carbon price movement (Alberola et al., 2008; Hintermann, 2010). When the study period is extended to Phase II, the impact of the energy industry on the carbon market becomes more difficult to measure (Wu et al., 2019). As the EU ETS is an emerging market, a continuous adjustment of policies take place during the market operation. The scholars have tried to better explain the operating mechanism of carbon prices from the perspective of market policies. Fan et al. (2017) and Guo et al. (2018) extract 50 policy announcements from the EU ETS and analysed their effects on carbon prices, revealing that only parts of the policies have influence on carbon price while the overall impact is small. Fan et al. (2016) and Wang et al. (2019) claim all external factors in the carbon market will be reflected in the trading behaviour of traders, whether these factors are energy prices or policy announcements. These factors affect the demand and supply of carbon allowances by influencing the traders' trading behaviour, which in turn affects the carbon price. These researchers found that compliance transactions will affect carbon price trends, while non-compliance transactions will affect carbon price fluctuations (Wang et al., 2019).

The above literatures study the causes of the carbon price fluctuations from the angle of micro behaviour. In order to fully explain the volatility of European carbon market, some scholars conduct their study from the macroeconomic perspective. In recent years, the economic policy uncertainty has attracted the attention of scholars (Brogaard and Detzel, 2015; Baker et al., 2016; Dai et al, 2021a). Among them, there is few literatures discussing the linkage between the economic policy uncertainty and the emerging carbon market. To make up for the shortcomings of the above-mentioned research about carbon market, this article explores the potential determinants of the volatility of carbon return from the view of the external economic-related policy environment. The carbon spot price in the EU ETS basically reflects the supply and demand of carbon allowances in the market, and the supply and demand of carbon allowances is a direct expression of enterprise production. Therefore, in theory, the fluctuation of the entire economic environment is the determinant of carbon price fluctuations. In terms of the theoretical framework between economic policy uncertainty and financial markets, there are voluminous studies continuously contributing to the literature. In particular, Ozsoylev and Werner (2011) constructed the theory to explain the role of compensated payoff under the ambiguous decision-making. This issue has been supported by the equilibrium model by Pástor and Veronesi (2013), implying the higher stock volatility under weak economy as uncertainties. In the same pipeline, Bialkowski et al. (2021) attempt to explain the academic puzzle by contributing two relevant factors to literature when the higher shake of economic policy uncertainties is not associated with volatility. It turns out that the low-quality information and divergence of attitudes could not link to the more volatile financial assets. From this starting point, we can form the research question whether the European carbon markets exhibit the higher or lower dynamics of risk under uncertainties. This is our basically fundamental concept for us to build up arguments based on findings and results.

In the current context of economic globalization, the European economy is an important part of the global economy. For long-term development performance, the European companies not only react to the European economic policies when making corporate decisions but pay attention to global economic policy. There we First determine the proxy variables for economic-related policy uncertain environment: European economic policy uncertainty and global economic policy uncertainty. Then, we detect the impact of realized volatility of the European carbon market on the long-term volatility of carbon spot return. This part serves as a reference for comparing the influences of economic policy uncertainty on carbon price returns. Finally, we analyse the impact of economic policy uncertainty on the volatility of the European carbon spot return. In the remainder of this introduction, we elaborate our contribution to the literature through three main points.

Firstly, this paper reveals the long-term impact of the realized carbon price return volatility on the carbon price return volatility. Second, this study uses the index of European economic policy uncertainty (EUEPU), according to the fluctuation of the entire European economic environment, as a proxy variable to analyse its impact on the carbon price return volatility in the EU ETS. Secondly, based on the background of economic globalization, the index of global economic policy uncertainty is used as another proxy variable to study the influence of economics on the EU ETS carbon price. Finally, by comparing the results of the above three models, the determinants of carbon price return volatility can be found, which can provide a valuable way of predicting the volatility of carbon price returns in the EU ETS and can provide policy implications for the operation of the carbon market.

The rest of the paper is organized as follows. Section 2 presents the data description. Section 3 introduces the methods. Section 4 shows the results of the empirical study. We elaborate our findings and the implications of the empirical results in Section 5. Section 6 concludes.

**2. Data Description**

As mentioned, the EU ETS has been running for three phases. Phase I (the pilot phase) was from 2005 to 2007; Phase II was from 2008 to 2012; and Phase III is from 2013 to 2020. All the three phases have been completed. According to the efficient market hypothesis (Fama, 1995), all the market information is reflected in the market asset price. In the paper, in order to learn about the European carbon market directly, we focus on the carbon spot market which attracts less attention. We manually capture the carbon spot price data from European Environmental Exchange (BlueNext), which is the largest spot market for carbon dioxide emission rights, and Intercontinental Exchange (ICE). Since BlueNext announced that it permanently closed the spot trade on December 5, 2012, we retrieve the remaining carbon spot price data from ICE official trading website. Limited to the computing equipment, we only extract the continue data from the starting trading day of carbon spot market (April 22, 2005) to September 15, 2015. Figure 1(a) displays the evolving trajectory of the European carbon spot price from April 22, 2005 to September 15, 2015. As illustrated in Figure 1(a), the carbon spot price is discontinuous across the Phase I to Phase II, dropping to approximately 0 euros by the end of Phase I. This phenomenon mainly results from the over-allocation of carbon allowances during the pilot period and the unavailability of the extra Phase I carbon allowances among Phase II. Moreover, the situation affects the efficiency of the market transactions. Therefore, the system of the European carbon market was adjusted and improved in many ways in Phase II to improve the efficiency of the market (Creti et al., 2012). One of the obvious adjustments of the rules of EU ETS is that the carbon allowances allocated in Phase II can be used in Phase III, which ensures the continuity of the carbon spot price across Phase II and Phase III. As shown in Figure 1(a), there is no break point in the second half of the carbon spot price curve. Since our research focus on the returns of the European carbon spot price, we do not take the pilot period into account in order to avoid facing the breaking point problem. In addition, the Phase II and Phase III of EU ETS are relatively mature, therefore we select the data of carbon spot price over the two phases in our research. That is, the sample period is January 1, 2008 to September 15, 2015.

Baker et al. (2016) construct a policy-related economic uncertainty index (EPU) for the United States, derived from newspapers, and they update the data regularly on the Economic Policy Uncertainty website (http://www.policyuncertainty.com/). The EPU index is the proxy variable for policy-related economic uncertainty. Many scholars study topics about policy uncertainty by using the proxy variable (Colombo, 2013; Brogaard and Detzel, 2015; Dai et al., 2019; Dai et al., 2021a). To measure European economic policy uncertainty, Baker et al. (2016) construct a European economic policy uncertainty index (EUEPU) based on the newspaper articles. They select five countries and then extract two newspapers per country for the EUEPU: Le Monde and Le Figaro for France, Handelsblatt and Frankfurter Allgemeine Zeitung for Germany, Corriere Della Sera and La Stampa for Italy, El Mundo and El Pais for Spain, and The Times of London and Financial Times for the United Kingdom. In our paper, we employ the EUEPU as the proxy variable to quantify the fluctuations in the European economic policy environment. Since more and more scholars have constructed different proxy indices for various countries, Dai et al. (2021b) figure out an index for the aggregate global economic policy uncertainty (GEPU) based on the principal component analysis (PCA). They calculate the PCA-based GEPU index by weighting the economic policy uncertainty indices from twenty economies, which are obtained from the economic policy uncertainty website (http://www.policyuncertainty.com). These twenty economies include Australia, Brazil, Canada, Chile, China, France, Germany, Greece, India, Ireland, Italy, Japan, Mexico, the Netherlands, Russia, South Korea, Spain, Sweden, the United Kingdom and the United States. In this paper, in order to match the sample period of European carbon spot price, we duplicate their work and obtain the GEPU time series from April 2005 to September 2015. We select this GEPU index as the proxy for fluctuations in the global economic environment (Dai et al, 2021). Figure 1 illustrates the evolving trend of our research objects, the European carbon spot price and the two economic policy uncertainties. From Table 1, we can see clearly that both EUEPU and GEPU rise sharply over the European debt crisis period. As a matter of fact, the valid data is between January 1, 2008 and September 15, 2015.

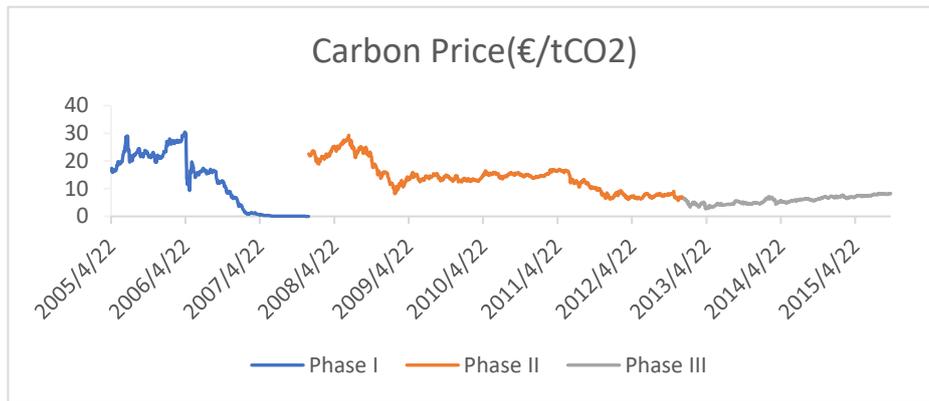

Figure 1(a). Carbon spot prices in EU ETS. Source: ICE and BlueNext.

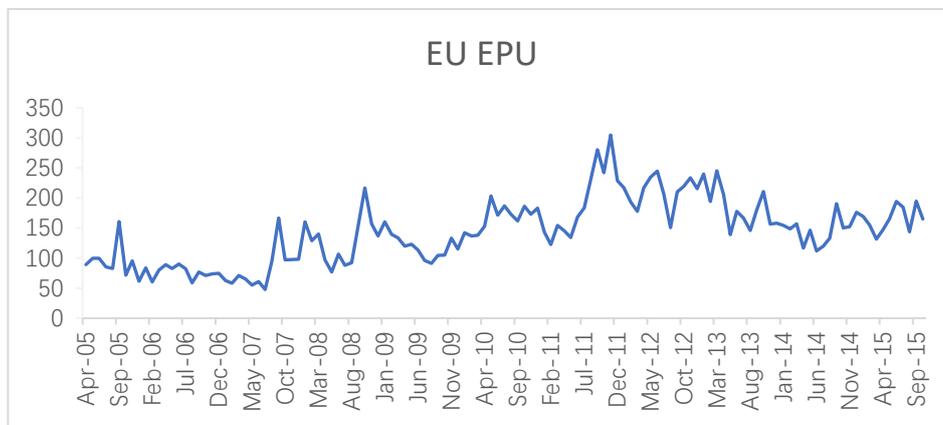

Figure 1(b). European economic policy uncertainty index

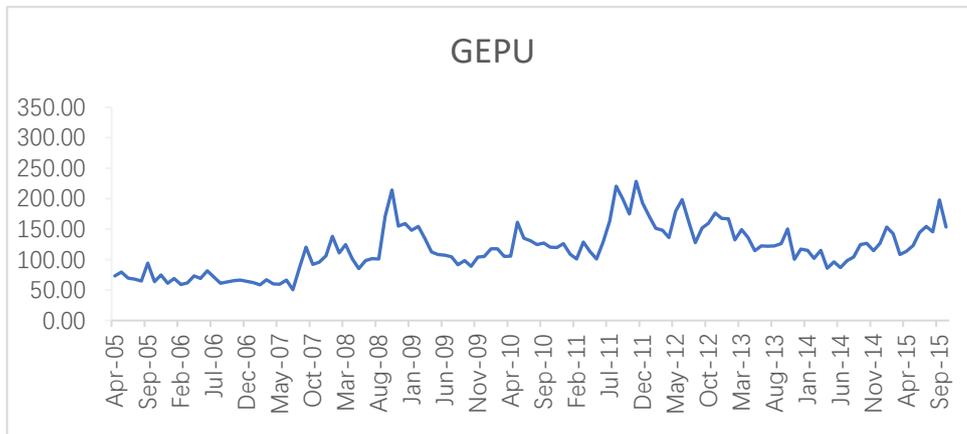

Figure 1(c). Global economic policy uncertainty index

Figure 1. The evolving trend of European carbon spot price and economic policy uncertainty.

In Table 1, we summarize the descriptive statistics of the natural logarithmic return series of the European carbon spot price (*RCP*), EUEPU and GEPU. According to the Jarque-Bera test, we find the carbon spot returns follow a normal distribution, as well as EUEPU and GEPU. Focusing on the ADF value, obtaining from unit root test, we find the carbon spot return time-series is stationary. When the statistical results for EUEPU and GEPU are compared, we find the average value of EUEPU is larger than that of GEPU. The maximum and median of EUEPU are both larger than that of GEPU, meanwhile, the standard deviation of EUEPU is larger than that of GEPU. That is, the economic policy uncertainty in Europe presents greater value and larger volatility than global uncertainty. Moreover, considering the results of the unit root test for EUEPU and GEPU, we find the two time-series are stationary, which implies these ample data can be used in the following GARCH model directly.

Table 1. Descriptive statistics of the variables

|  | RCP | EUEPU | GEPU |
|---|---|---|---|
| Mean | − 0.0003 | 143.4809 | 117.1269 |
| Med. | 0.0000 | 145.5086 | 114.6188 |
| Max. | 7.7231 | 304.6002 | 228.3534 |
| Min. | − 1.3863 | 47.6923 | 50.5451 |
| Std. Dev. | 0.1610 | 53.8257 | 38.9355 |
| Skew. | 40.8579 | 0.3411 | 0.4895 |
| Kurt. | 70.6725 | 2.6335 | 2.8979 |
| J.-B. | 1977.5350*** | 3.1739** | 5.1273* |
| ADF | − 15.6923*** | − 3.5247*** | − 3.2958** |

***, **, * denotes rejection of null of normal distribution or stationarity at 1%, 5%, 10% significance level respectively.

**3. Methodology**

*3.1 GARCH-MIDAS model*

There are various approaches to modelling and predicting carbon market volatility (Viteva et al., 2014; Dutta, 2018; Wang et al., 2020). Among them, the GARCH-class models occupy an important position. Most empirical research using GARCH-class models is based on the same low-frequency data (Wang et al., 2020). To improve these methods based on the same frequent data, Ghysels et al. (2004, 2007) propose a mixed data sample (MIDAS) regression model which eliminates the restriction of the same frequent data. With respect to the volatility research, the traditional GARCH-class models can be conducted successfully which also needs the same frequent data. Engle et al. (2008) introduce the spine-GARCH model to capture both short-term and long-term dynamic behaviour of market volatility. Among them, the long-term dynamic behaviour is described as a combination of macroeconomic effects. Engle et al. (2013) establish the GARCH-MIDAS model by integrating the MIDAS regression with the common unit GARCH (1,1) model. They employ the model to investigate the contribution of macroeconomic variables to explain and predicting the long-term volatility of the stock market. In addition, in their paper, spine-GARCH model and the GARCH-MIDAS model are compared, the empirical results show that the GARCH-MIDAS model performs better. Afterwards, many scholars have studied the effects of macro low-frequency variables on micro high-frequency variables (Asgharian et al., 2013; Wei et al., 2017; Fang et al., 2018). For instance, Wei et al. (2017) examine the influence of oil demand, supply and economic policy uncertainty on the volatility of crude oil spot market using the GARCH-MIDAS model. Fang et al. (2018) apply the GARCH-MIDAS model to explore whether global economic policy uncertainty could forecast the volatility of gold futures returns. It is worth mentioning that the global economic policy uncertainty index appearing in Fang et al. (2018) is a GDP-weighted average of 16 economies' EPU indices, which is constructed by Davis (2016). Dai at al. (2021) adopt principle component analysis to construct an alternative index for the aggregate GEPU, which is concise and effective. Furthermore, Dai at al. (2021) compare the global index (Davis, 2016) with PCA-based GEPU. Their findings show that the PCA-based GEPU preforms better, which is the reason we select it as the proxy variable for global economic policy uncertainty in Section 2. Since there is little research explore the link between economic uncertainty and the carbon emission allowance trading system, in this paper, we employ the GARCH-MIDAS model to investigate whether the European and global policy-related economic uncertainty have an impact on the long-term volatility of the European carbon spot market.

The GARCH-MIDAS model combines the high-frequency micro variables and low-frequency macro variables. It decomposes the volatility of low-frequency time series into two components: short-term volatility and long-term volatility. The long-run volatility is determined by low-frequency macro factors, while the short-run volatility depends on the dynamics of the high-frequency time series. In this paper, we utilize the GARCH-MIDAS model to investigate the impact of EUEPU and GEPU on the volatility of the European carbon spot returns. The GARCH-MIDAS model is formally expressed as follows. The carbon spot return on day $i$ in period $t$, $r_{i,t}$, obeys the following process:

$$r_{i,t} - E_{i-1,t}(r_{i,t}) = \sqrt{\tau_t g_{i,t}} \cdot \varepsilon_{i,t} , \qquad \forall i = 1,2,\cdots,N_t \tag{1}$$

$$E_{i-1,t}(r_{i,t}) = \mu \tag{2}$$

$$\varepsilon_{i,t} | \Phi_{i-1,t} \sim N(0,1) \tag{3}$$

where $\Phi_{i-1,t}$ is the information set up to day $i-1$, $\varepsilon_{i,t}$ is the innovation term and $\mu$ represents the expectation of carbon spot return at day $i$ in period $t$. Following Wei et al. (2017), we set $\mu$ as a constant, since the mean daily return of the carbon spot is quite small. Equation (1) implies that the volatility of the return is decomposed into two parts: the short-term volatility component, represented by $g_{i,t}$, and the long-term volatility component, represented by $\tau_t$.

The dynamics of the short-term volatility component are set to be a common GARCH (1,1) process:

$$g_{i,t} = (1 - \alpha - \beta) + \alpha \frac{(r_{i-1,t} - \mu)^2}{\tau_t} + \beta g_{i-1,t} \tag{4}$$

where $\alpha > 0$, $\beta > 0, \alpha + \beta < 1$. $\beta$ denotes the degree of volatility clustering (the larger the $\beta$, the stronger the volatility clustering). On the other side, the long-term volatility component, $\tau_t$ is specified as smoothed realized volatility via MIDAS regression:

$$\tau_t = m + \theta \sum_{k=1}^{K} \varphi_k RV_{t-k} \tag{5}$$

$$RV_t = \sum_{i=1}^{N_t} r_{i,t}^2 \tag{6}$$

where $K$ is the number of periods over which we smooth the realized volatility, $m$ is the intercept term of the MIDAS regression. $RV$ is the realized volatility of the European carbon spot returns, and $\theta$ is the slope which denotes the effect of the realized volatility on the long-term volatility of carbon spot return. In Equation (6), $r_{i,t}$ represents the daily return on day $i$ in period $t$; $N_t$ means that each period has $N_t$ trading days. The data we use to calculate the realized volatility is high-frequency (daily data), while the macro-economic variable in this paper is the monthly data, low-frequency data. In order to obtain RV, we calculate the cumulative sum of the square of carbon's daily return over a fixed period (a trading month, $N_t = 22$). In our paper, $r_{i,t}$ is the European carbon spot's daily return which is the available data. Following Engle and Rangel (2008), the weighting scheme in Equation (5) is assigned by a two-parameter Beta polynomial:

$$\varphi_k(\omega_1, \omega_2) = \frac{(k/K)^{\omega_1 - 1}(1 - k/K)^{\omega_2 - 1}}{\sum_{j=1}^{K}(k/K)^{\omega_1 - 1}(1 - j/K)^{\omega_2 - 1}} \tag{7}$$

Always we pose $\omega_1 = 1$ (Asgharian et al., 2013), and then the two-parameter Beta polynomials degrade to a one-parameter Beta polynomial:

$$\varphi_k(\omega) = \frac{(1 - k/K)^{\omega - 1}}{\sum_{j=1}^{K}(1 - j/K)^{\omega - 1}} \tag{8}$$

Combining Equations (1), (4) and (5), we obtain the GARCH-MIDAS model for the realized volatility of European carbon spot return (GARCH-MIDAS-RV).

Next, with respect to the European carbon market, we turn to the GARCH-MIDAS model which directly incorporates the external macroeconomic factor, EUEPU and GEPU. For EUEPU, the corresponding long-term volatility term $\tau_t$ is expressed as follows:

$$\tau_t = m + \theta \sum_{k=1}^{K} \varphi_k EUEPU_{t-k} \qquad (9)$$

where the slope $\theta$ denotes the impact of EUEPU on the long-term volatility of the European carbon spot return. Combining Equations (1), (4) and (9), we get the GARCH-MIDAS model for EUEPU (GARCH-MIDAS-EUEPU). In terms of GEPU, the corresponding long-term volatility term $\tau_t$ has the following expression:

$$\tau_t = m + \theta \sum_{k=1}^{K} \varphi_k GEPU_{t-k} \qquad (10)$$

where the slope $\theta$ implies the impact of GEPU on the long-term volatility of the European carbon spot return. All weighting schemes $\varphi_k$ appearing in the long-term volatility specifications (9) and (10) have the same definition, which is given by Equation (8).

Finally, the total conditional variance is defined as the following specification:

$$\sigma_{i,t}^2 = \tau_t \cdot g_{i,t} \qquad (11)$$

*3.2 Model evaluation*

We calibrate the GARCH-MIDAS models in Section 3.1 using full sample data from January 1, 2008 to September 15, 2015. The calibrations could assist to analyse the explanatory ability of the models, which indicates the impact of macro variables on long-term volatility of European carbon spot return over the full sample period. Furthermore, in order to make the model evaluation, we divide the full ample into two parts, the in-sample part and the out-of-sample part. Firstly, we estimate the models using the in-sample data, and then apply the estimated parameters to out-of-sample variance prediction. In our empirical work, we select the out-of-sample period from October 1, 2014 to September 15, 2015. As for the rest sample (in-sample) data, we choose a five-year estimation window, which is from January 2010 to October 2014. As equation (5) required, *K* lagged periods are needed for calculating the current total influence of the macro factor, since the influence is persistent and time-varying. In our model, we set *K*=24 (Asgharian et al., 2013), which means two years' lagged data before the estimation window is needed to calculate the historical realized volatility of the European carbon spot return and the economic policy uncertainty from Europe and global scale. That is, we use the data from January 2008 to December 2009 to estimate the aggregate impact of the realized volatility of the European carbon spot return and economic policy uncertainty in January 2010. Finally, four popular loss functions, which are the root mean squared error (*RMSE*), root mean absolute error (*RMAE*), the root mean squared deviation (*RMSD*), and root mean absolute deviation (*RMAD*), are employed to perform the model evaluation. The four loss functions are defined as follows:

$$RMSE = \sqrt{\frac{1}{T}\sum_{t=1}^{T}(\sigma_{t+1}^2 - E_t(\sigma_{t+1}^2))^2} \qquad (11)$$

$$RMAE = \sqrt{\frac{1}{T}\sum_{t=1}^{T}|\sigma_{t+1}^2 - E_t(\sigma_{t+1}^2)|} \qquad (12)$$

$$RMSD = \sqrt{\frac{1}{T}\sum_{t=1}^{T}(\sigma_{t+1} - E_t(\sigma_{t+1}))^2} \qquad (13)$$

$$RMAD = \sqrt{\frac{1}{T}\sum_{t=1}^{T}|\sigma_{t+1} - E_t(\sigma_{t+1})|} \qquad (14)$$

where $\sigma_{t+1}^2$ is the actual daily total variance at $t+1$, $E_t(\sigma_{t+1}^2)$ is the predicative daily total variance for $t+1$ and $T$ is the length of the forecasting interval. The loss functions of the full sample for the model characterize the model estimation error. Moreover, the loss functions of out-of-sample for the model describe the prediction error of the model.

Finally, for the sake of comparing the predictive accuracy of two competing models directly, DM test proposed by Diebold and Mariano (1995) is employed:

$$DM = \frac{\bar{D}}{\sqrt{var(D_t)}} \sim N(0,1) \qquad (15)$$

$$D_t = e_{A,t}^2 - e_{B,t}^2 \qquad (16)$$

where $e_{A,t}$ and $e_{B,t}$ are the predictive error of two rival models A and B respectively, while $\bar{D}$ is the mean of the time-series $(D_t)$, and $var(D_t)$ is the variance of $D_t$.

## 4. Empirical results

*4.1 Results from GARCH-MIDAS-RV model*

In this section, we calibrate the GARCH-MIDAS-RV model using the full sample data. The in-sample estimation can be viewed as the robust test for the full-sample calibration of the model. the in-sample parameters of the model. In order to reserve one-year time interval, September 16, 2014 to September 15, 2015, for testing the forecasting ability of the model, the data for the European carbon spot return between January 1, 2008 and September 15, 2014 is selected for the in-sample estimation of the GARCH-MIDAS-RV model. The volatility of the carbon spot return in last year of the full sample is predicted based on the in-sample estimation results. In Table 2, we present the results of the full-sample calibration and in-sample estimation, including all the parameters in the model and the corresponding Akaike information criterion (AIC). As we can see from Table 2, apart from $\mu$, all the other coefficients are significant at least 5% level, which indicates the GARCH-MIDAS-RV model is constructed rationally and validly. As there is no essential difference in the parameters' estimation from the two samples period, the results of in-sample estimation show the model is we use the data for the different sample period. According to the value of $\beta$, we find the short-term volatility of the European carbon spot return is highly aggregated, which is consistent with the findings of Guo et al. (2020). Due to the allocation and surrender mechanism in the EU ETS, transactions are concentrated into these periods (Wang et al., 2020). Focusing on the most important coefficient $\theta$, we find the values are positive significantly both in full ample and in-sample situation, which implies the realized volatility of the European carbon spot return produces positive influence on the long-term volatility of the European carbon spot return itself. The long-term volatility of the European carbon spot return will be heavy when its realized volatility is heavy. While the decline of the realized volatility will weaken the long-term volatility of the European carbon spot return.

Table 2. Estimated results of GARCH-MIDAS-RV model

| Para. | Val. | Std. | t stat. | Val. | Std. | t stat. |
|---|---|---|---|---|---|---|
| | | Full sample | | | In-sample | |
| $\mu$ | 0.0006 | 0.0005 | 1.21 | 0.0002 | 0.0007 | 0.34 |
| $\alpha$ | 0.1221*** | 0.0109 | 11.18 | 0.1748*** | 0.0234 | 7.46 |
| $\beta$ | 0.8608*** | 0.0222 | 66.43 | 0.7190*** | 0.0509 | 14.13 |
| $\theta$ | 0.1855*** | 0.0218 | 8.34 | 0.0428*** | 0.0037 | 11.44 |
| $\omega$ | 2.8589** | 1.1089 | 2.58 | 45.8130*** | 14.6270 | 3.13 |
| $m$ | 0.0184*** | 0.0037 | 5.02 | 0.0002*** | $5.0731 \times 10^{-5}$ | 4.14 |
| AIC | | 7.4962 | | | 7.6625 | |

Note: Val. denotes the value of the corresponding parameter, Std. is the standard error and t stat. is the t statistics. AIC is the Akaike information criterion, a measure of the statistical model goodness of fit. ***, **, * denote statistical significance at 1%, 5%, 10 % level respectively.

Table 3 reports the results of the four loss functions for the GARCH-MIDAS-RV model. All values of *RMSE*, *RMAE*, *RMSD* and *RMAD* are very small, indicating that the model is relatively accurate in predicting the volatility of the European carbon spot return. According to the results of this model, the EU ETS transitioned in a relatively stable fashion from Phase II to Phase III, although the market mechanism needed some fine-tuning (Guo et al., 2019). The coverage of the included industries is more complete; for example, the aviation industry is also included as a large-emission industry (Fan et al., 2017). The stability of the EU ETS is not only reflected in the continuity of the carbon spot price (Figure 3), but also in the volatility dynamics of the carbon spot return. The long-term volatility of the European carbon spot return can be predicted based on the volatility patterns of the foregoing carbon price return. That is, from the perspective of its own dynamics, the finding provides guidance for the policy-makers in the EU ETS to avoid the drastic fluctuations of the European carbon market on a long-run time scale.

Table 3. Evaluation results of GARCH-MIDAS-RV model

| Sample period | *RMSE* | *RMSD* | *RMAE* | *RMAD* |
|---|---|---|---|---|
| Full sample | $4.984 \times 10^{-3}$ | $2.469 \times 10^{-2}$ | $3.501 \times 10^{-2}$ | 0.1331 |
| Out-of-sample | $7.583 \times 10^{-4}$ | $1.443 \times 10^{-2}$ | $2.014 \times 10^{-2}$ | 0.1078 |

*4.2 Results from GARCH-MIDAS-EUEPU model*

The GARCH-MIDAS-EUEPU model can connect the long-term volatility of the European carbon spot return with the European economic policy uncertainty. Table 4 reports the parameter estimations under circumstance of full-sample and in-sample data provided. Almost all the parameter estimations are significant at the 10% level, which means the GARCH-MIDAS-EUEPU model are constructed rationally and validly. Comparing the values of $\alpha$ and $\beta$ in Table 2 and Table 4, we find the short-term volatility of the European carbon spot return estimated from the GARCH-MIDAS-RV model and GARCH-MIDAS-EUEPU model can be depicted similarly. This finding indicates the macro factor indeed do not have significant effect on the short-term volatility, which is mainly determined by the internal dynamins of the carbon spot return series. The value of $\theta$ is a positive number, indicating that the uncertainty of the European economic policy produces positive impact on the long-term volatility of the carbon spot return. That is, the increased fluctuation of European policy environment will aggravate the volatility of the carbon spot return. Comparing the AIC of the GARCH-MIDAS-RV model (7.4962) and the GARCH-MIDAS-EUEPU model (7.5017) in the case of full sample, we find the GARCH-MIDAS-RV model hold the stronger explanatory ability of describing the volatility pattern of the European carbon spot return. However, the external policy-related factor still plays the similar role as the internal time-series dynamics in explaining the volatility of European carbon spot return.

Table 4. Estimated results of GARCH-MIDAS-EUEPU model

| Para. | Val. | Std. | t stat. | Val. | Std. | t stat. |
|---|---|---|---|---|---|---|
| | | Full sample | | | In-sample | |
| $\mu$ | 0.0006 | 0.0005 | 1.12 | $-0.0039^{***}$ | 0.0005 | $-7.79$ |
| $\alpha$ | $0.1233^{***}$ | 0.0111 | 11.07 | $0.0715^{***}$ | 0.0044 | 16.07 |
| $\beta$ | $0.8702^{***}$ | 0.0117 | 74.38 | $0.8177^{***}$ | 0.0132 | 61.89 |
| $\theta$ | $0.0059^{*}$ | 0.0035 | 1.75 | $0.0018^{***}$ | $8.0514\times10^{-5}$ | 22.43 |
| $\omega$ | $1.5033^{***}$ | 0.5378 | 2.79 | $5.0341^{***}$ | 0.8182 | 6.15 |
| $m$ | $-0.0007$ | 0.0007 | $-0.98$ | $-0.0003^{***}$ | $1.3417\times10^{-5}$ | $-19.87$ |
| AIC | | 7.5017 | | | 7.80378 | |

Note: Val. denotes the value of the corresponding parameter, Std. is the standard error and t stat. is the t statistics. AIC is the Akaike information criterion, a measure of the statistical model goodness of fit. $^{***}$, $^{**}$, $^{*}$ denote statistical significance at 1%, 5%, 10 % level respectively.

The values of the four loss functions for the GARCH-MIDAS-EUEPU model, not only the out-of-sample forecasting loss but also the full sample estimation loss, are shown in Table 5. All the values of the four loss functions for the out-of-sample forecasting are also relatively small, indicating that the volatility of carbon spot return can be predicted reasonably by the GARCH-MIDAS-EUEPU model. This finding shows that the external variable EUEPU can be viewed as the volatility factor for the European carbon spot market. Comparing the values of the four loss functions in Table 3 and Table 5, we discover that the values of *RMSE*, *RMSD*, *RMAE*, *RMAD* from the GARCH-MIDAS-RV model are smaller than that from the GARCH-MIDAS-EUEPU model respectively, which indicates the GARCH-MIDAS-RV model performs better both in the full-sample estimation and the out-of-sample forecasting. In term of the full sample, the results from the model evaluation are consistent with that from the model estimation. Since the EU ETS is not only affected by the European internal policy, but also influenced by other countries' policy for carbon emission, to better simulate the volatility of the European carbon spot return, the GARCH-MIDAS-GEPU is constructed based on global economic policy uncertainty, as discussed in Section 4.3.

Table 5. Evaluation results of GARCH-MIDAS-EUEPU model

|  | RMSE | RMSD | RMAE | RMAD |
|---|---|---|---|---|
| Full sample | $5.052\times10^{-3}$ | $2.820\times10^{-2}$ | $3.879\times10^{-2}$ | 0.1452 |
| Out-of-sample | $7.743\times10^{-4}$ | $1.481\times10^{-2}$ | $2.112\times10^{-2}$ | 0.1120 |

*4.3 Results of GARCH-MIDAS-GEPU model*

The GARCH-MIDAS-GEPU model is employed to analyse the effect of the global economic policy uncertainty on the volatility of European carbon spot return. Furthermore, the empirical results will enhance our comprehension to the linkage between the external macroeconomic factor and European carbon market. The calculated results of the full-sample calibration and the in-sample estimation from the GARCH-MIDAS-GEPU model are shown in Table 6. Since almost all the estimated coefficients are significant at the least 5% level except $\mu$, the GARCH-MIDAS-GEPU model are constructed rationally and validly. In addition, the pairwise estimated coefficients of different samples keep the same sign, hence the model passes the robustness test. The estimations of $\alpha$ and $\beta$ from the three GARCH-MIDAS (RV, EUEPU, GEPU) models have little difference, 0.1221 and 0.8608, 0.1233 and 0.8702，0.1171 and 0.8643 respectively. The GARCH-MIDAS-GEPU model play almost the same role in extracting the short-term volatility from the total volatility of the European carbon spot market as the previous two models. The estimations of $\theta$ are positive both in full-sample and in-sample case, indicating that GEPU has positive effect on the long-term volatility of the carbon spot market in the EU ETS. Comparing the value of $\theta$ in Table 6 and Table 4, we find GEPU (0.0061) and EUEPU (0.0059) cause the similar (positive) influence on the volatility of the European carbon market. The extra-European uncertainty indeed produces a certain impact on volatility of the European carbon spot return. Numerically, GEPU has larger impact on the long-term volatility than EUEPU, that is, a unit increased standard deviation in GEPU will bring more change in the long-term volatility of the European carbon market than EUEPU. Considering the numerical value of GEPU is smaller and with lower volatility than that of EUEPU generally (see Figure 1), the change in the external policy uncertain environment leads to the increase of GEPU would be harder than EUEPU. The values of AIC in Table 6 (7.4955) and Table 4 (7.5017) show the explanatory ability of the GARCH-MIDAS-GEPU model is stronger than that of the GARCH-MIDAS-EUEPU model. Comparing the values of AIC in Table 6 and Table 2, we find the GARCH-MIDAS-GEPU model even performs better in describing the volatility of the European carbon market than its realized volatility.

Table 6. Estimated results of GARCH-MIDAS-GEPU model

| Para. | Val. | Std. | t stat. | Val. | Std. | t stat. |
|---|---|---|---|---|---|---|
| | Full sample | | | In-sample | | |
| $\mu$ | 0.0007 | 0.0005 | 1.32 | 0.0003 | 0.0003 | 0.45 |
| $\alpha$ | 0.1171*** | 0.0108 | 10.88 | 0.1388*** | 0.0126 | 10.99 |
| $\beta$ | 0.8643*** | 0.0129 | 67.08 | 0.8383*** | 0.0151 | 55.63 |
| $\theta$ | 0.0061*** | 0.0013 | 4.748 | 0.0071*** | 0.0015 | 4.76 |
| $\omega$ | 1.2256*** | 0.1920 | 6.38 | 1.4225*** | 0.2080 | 6.84 |
| $m$ | − 0.0004** | 0.0002 | − 2.06 | − 0.0005** | 0.0002 | − 2.41 |
| AIC | | 7.4955 | | | 7.6479 | |

Note: Val. denotes the value of the corresponding parameter, Std. is the standard error and t stat. is the t statistics. AIC is the Akaike information criterion, a measure of the statistical model goodness of fit. ***, **, * denote statistical significance at 1%, 5%, 10 % level respectively.

As for the model evaluation, Table 7 reports the results of four loss functions for the GARCH-MIDAS-GEPU model. Comparing the results of loss functions in Table 5 and Table 7, the values of *RMSE*, *RMAE*, *RMSD* and *RMAD* from the GARCH-MIDAS-GEPU model are all much smaller than that from GARCH-MIDAS-EUEPU model. Therefore, in terms of the full sample, the results are consistent with the conclusion related to AIC. The GARCH-MIDAS-GEPU model has stronger explanatory ability in a whole time period. With respect to the out-of-sample analysis, the volatility of European carbon spot return can be better predicted based on GARCH-MIDAS-GEPU model than on GARCH-MIDAS-EUEPU model. Analysing the results of model evaluation in Table 3, Table 5 and Table 7, the four loss functions for the GARCH-MIDAS-GEPU model all have the smallest value, which implies that amongst these three factors (RV, EUEPU, GEPU), GEPU is the best volatility predictor of the European carbon spot market.

Table 7. Evaluation results of GARCH-MIDAS-GEPU model

| | *RMSE* | *RMSD* | *RMAE* | *RMAD* |
|---|---|---|---|---|
| Full sample | 4.956×10$^{-3}$ | 2.460×10$^{-2}$ | 3.488×10$^{-2}$ | 0.1334 |
| Out-of-sample | 7.472×10$^{-4}$ | 1.393×10$^{-2}$ | 1.962×10$^{-2}$ | 0.1056 |

With respect to the DM test, Table 8 reports the calculated results between any two competing models. The results further suggest that GEPU is the best volatility predictor of the European carbon spot market among these three factors, which is consistent with the results of loss functions. The global uncertain policy environment would affect the volatility of European carbon spot market beyond European uncertain policy environment.

Table 8. Diebold and Mariano test (1995) results for the competing models

| Model | GARCH-MIDAS-RV | GARCH-MIDAS-EUEPU | GARCH-MIDAS-GEPU |
|---|---|---|---|
| GARCH-MIDAS-RV | / / | −0.0652** (−2.7354) | 0.1652*** (6.9260) |
| GARCH-MIDAS-EUEPU | 0.0652** (2.7354) | / / | 0.0823*** (3.4525) |
| GARCH-MIDAS-GEPU | −0.1652*** (−6.9260) | −0.0823*** (−3.4525) | / / |

Note: This table reports the Diebold and Mariano test (1995) results between any two models among all the three models (GARCH-MIDAS-RV/EUEPU/GEPU). The out-of-sample for DM test is from September 16, 2014 to September 15, 2015. " / " represents that the row model cannot be compared with the corresponding column model. "−" denotes the row model is more accurate than the corresponding column model; "+" is the other way around. The numbers in the parentheses are the t statistic. ***, **, * denote significance at 1%, 5%, 10% level respectively.

## 5. Discussion

The empirical results show that the estimated short-term volatility of the European carbon spot market from the three GARCH-MIDAS models have strong volatility clustering, which is consistent with Guo et al. (2020). In their paper, they also point out that transactions are concentrated into these periods of volatility clustering, due to the allocation and surrender mechanism in the EU ETS. With regard to the estimated long-term volatility of the European carbon spot market, we find both the internal and external factors are the determinants. The results from Section 4.1 indicate the realized volatility of European carbon market as an internal factor can be used to illustrate and forecast the volatility of European carbon market itself. There are other literature exploring the internal determinants for the carbon market volatility (Viteva et al., 2014; Wang et al, 2020). Viteva et al. (2014) focus on the European carbon options and its total volatility. They detect the forecasting accuracy of the implied volatility of options on futures contracts for the delivery of $CO_2$ emission allowances. Our article supplements the evidence for the forecasting effect of internal dynamics on the volatility of European carbon market. In addition, compared with other mature financial markets (Engle et al., 2013; Wei et al., 2017), the European carbon market also presents the phenomenon of mature market. The findings can be used as the guidance for the industries in EU ETS, the carbon market participants and the government policy-makers.

On the other hand, lots of literature investigate whether the external economic variables can be the determinants for the volatility of the European carbon market (Zhang and Sun, 2016; Kara et al., 2008; Dutta, 2018). Dutta (2018) examines the influence of oil market uncertainty on the emission price volatility by using the crude oil volatility index. Our paper documents that the external macro aggregate policy uncertainty in different spatial scale have positive impact on the volatility of the European carbon spot market. The findings show the direct statistical evidence that there is linkage between the macroeconomic variable and European carbon market. Moreover, the extra-European information about the policy can take effect on the European carbon market, which is consistent with the existing literature (Calel and Dechezlepretre, 2016). According our analysis, these findings have the reference significance for the industries in EU ETS, the carbon market participants and the government policy-makers. They should take the changeable policy environment, not only within Europe but global scale, into full account when they make the trading or policy decisions.

Besides the discussion of intuitive implication about statistical results, we explore the latent mechanisms through which economic uncertainty affects the carbon emission trading system in this section. Fan et al. (2016) divide the trading behaviours in the EU ETS into three categories: compliance trading behaviours of emitting companies for reducing their positions, non-compliance trading behaviours for increasing the positions of emitting companies, and the speculative trading behaviours of financial intermediaries. Among these, the non-compliance trading behaviours of emitting companies and the speculation trading behaviours of financial intermediaries will aggravate the volatility of carbon market. According to the results from the GARCH-MIDAS-EUEPU model and the GARCH-MIDAS-GEPU model, we find the phenomenon in Wang et al. (2020) can be fundamentally explained from the perspective of macroeconomic influences. First, the economic policy uncertainty will cause enterprises to continuously adjust their expectations for production capacity, so they will engage in non-compliance trading, which will affect the volatility of carbon market return. Second, the economic policy uncertainty provides a speculative environment for professional financial intermediaries, which exacerbates the information asymmetry between emitting companies and financial intermediaries. Thereupon, the volatility of carbon market return will increase.

Our findings are inclined to the ubiquitous evidence that the higher economic policy uncertainties correlate with the long-term volatility (Kelly et al., 2016). Our study also contributes a characteristic of carbon trading market, implying the convergence of investors' attitudes to impede the environmental degradation in European countries while the US presidents do not come into agreement to tackle environmental issues. For example, the policies of Trump loosening the heavy constraints of carbon emission would benefit for some industries (Nerger et al., 2021). By doing this, the market might experience beneficial imbalances, implying the possibility of higher uncertainties. However, the unified attitudes towards environment in EU could not be considered as the mitigating mediator in alleviating volatility. Thus, our study sheds a new light that economic policy uncertainties could drive higher shocks in carbon trading markets in the long run. Concomitantly, the current literature also indicate the investment behaviour in the EU zones with a financial accelerator mechanism (Strobel et al., 2018). Overall, data from different estimations in our study draw the same picture that behaviour under higher uncertainties requires the risk premium for compensating what the investors are bearing, which has been mentioned in the current literature (Pástor and Veronesi, 2013).

## 6. Conclusion

This paper studies for the first time the impact of economic policy uncertainty on the volatility of carbon spot return in the EU ETS, exploring the fundamental determinants of the European carbon spot return volatility. In this study, we construct the GARCH-MIDAS-RV model to quantify the influence of the realized volatility on the volatility of European carbon spot return. The results from the GARCH-MIDAS-RV model can be seen as a reference for comparison with the other GARCH-MIDAS models set up in this paper. Then, the GARCH-MIDAS-EUEPU model and the GARCH-MIDAS-GEPU model are constructed to analyse the impact of the European economic policy uncertainty and global economic policy uncertainty on the volatility of European carbon spot return. Furthermore, the predictive ability to the volatility of European carbon spot return is investigated according to the in-sample estimated results of all the three GARCH-MIDAS models.

Through comparing the performance of the estimation and prediction from the three models, this paper highlights three main findings which are listed as follows. Firstly, the empirical results of GARCH-MIDAS-RV model indicate the internal factor, the realized volatility, can be viewed as a determinant of the long-term volatility of the European carbon spot market. In addition, the realized volatility holds certain forecasting power for the future volatility of the European carbon spot market Secondly, both the EUEPU and GEPU are the external determinants which can be used to describe the long-term volatility of the European carbon spot return. The impact of GEPU on the long-term volatility of the European carbon spot return is stronger than EUEPU when the same change in the two economic policy uncertainty occur. Compared with other models, the GARCH-MIDAS-GEPU is the best model to characterize the volatility of the European carbon market. Thirdly, the empirical results show that compared with RV and EUEPU, GEPU is the best predictor for the volatility of the European carbon spot return. The extra-European policy uncertainty can also take effect on the European carbon market. The probable reason is that, in the current context of global economic integration, European companies will revise their expectations of the European future economy based on the global economic environment, and adjust their demand for carbon allowances in the EU ETS.

The research results of this article can provide some valuable implications for the development and improvement of the EU ETS. First, market managers can make predictions for the volatility of European carbon market based on the European and global economic policy uncertainty, which contributes to formulating the corresponding market policy. The valid predictivity can prevent possible market turmoil by stabilizing market traders' sentiment. Secondly, market traders can also forecast the European carbon market volatility according to these models, especially the emitting companies needing to engage in transactions to achieve compliance in the EU ETS. Thus, they can lock in the carbon price before the high risk occurs by purchasing corresponding futures of carbon allowance. Overall, both the market participants and the policy-makers can control the risk from the European carbon market based on the prediction for its volatility.

This paper provides a new perspective to understand the volatility of European carbon market. However, there are also some shortcomings. As a matter of fact, the European countries hold different levels of participation in the EU ETS. Some countries, such as Germany and the United Kingdom, are more active in market transactions than other countries. Therefore, the EU ETS may be more sensitive to the economic policy uncertainty from these kinds of countries. Limited to the data, we cannot investigate the different impact of the economic policy uncertainty from various countries in the EU ETS. This work will be the direction of our future research.

**Acknowledgment**

This is a project funded by China Postdoctoral Science Foundation (Grant No.: 2021M691015). This work was also supported by National Natural Science Foundation of China (Grants Nos. U1811462, 71532009, 71671180 and 71804177). This research is also partially funded by University of Economics Ho Chi Minh City (Vietnam) for the third author.